\documentclass[aps,10pt,prb,twocolumn,letterpaper,superscriptaddress]{revtex4}

\usepackage{amsmath}
\usepackage{epsfig}
\usepackage{graphicx, caption}
\usepackage{latexsym}
\usepackage{amssymb}
\usepackage{color}
\usepackage{amsfonts}
\usepackage{bm}
\usepackage{upgreek}

\begin{document}
\title{Persistent low-energy phonon broadening near the charge order \textit{q}-vector in bilayer cuprate Bi$_2$Sr$_2$CaCu$_2$O$_{8+\delta}$}

\author{Y.~He}
%\email[]{yuhe@stanford.edu}
\affiliation{Department of Applied Physics, Stanford University, Stanford, California 94305, USA}
\affiliation{Stanford Institute for Materials and Energy Sciences, SLAC National Accelerator Laboratory, 2575 Sand Hill Road, Menlo Park, California 94025, USA}

\author{S.~Wu}
\affiliation{Department of Physics, University of California, Berkeley, California 94720, USA}

\author{Y.~Song}
\affiliation{Department of Physics, University of California, Berkeley, California 94720, USA}

\author{W.-S. ~Lee}
\affiliation{Stanford Institute for Materials and Energy Sciences, SLAC National Accelerator Laboratory, 2575 Sand Hill Road, Menlo Park, California 94025, USA}

\author{A. H.~Said}
\affiliation{Advanced Photon Source, Argonne National Laboratory, Argonne, Illinois 60439, USA}

\author{A.~Alatas}
\affiliation{Advanced Photon Source, Argonne National Laboratory, Argonne, Illinois 60439, USA}

\author{A.~Bosak}
\affiliation{European Synchrotron Radiation Facility, BP 220, F-38043 Grenoble Cedex, France}

\author{A.~Girard}
\affiliation{European Synchrotron Radiation Facility, BP 220, F-38043 Grenoble Cedex, France}

\author{S. M.~Souliou}
\affiliation{European Synchrotron Radiation Facility, BP 220, F-38043 Grenoble Cedex, France}

\author{A.~Ruiz}
\affiliation{Department of Physics, University of California, Berkeley, California 94720, USA}

\author{M.~Hepting}
\affiliation{Stanford Institute for Materials and Energy Sciences, SLAC National Accelerator Laboratory, 2575 Sand Hill Road, Menlo Park, California 94025, USA}

\author{M.~Bluschke}
\affiliation{Max Planck Institute for Solid State Research, Heisenbergstr. 1, 70569 Stuttgart, Germany}
\affiliation{Helmholtz-Zentrum Berlin f\"{u}r Materialien und Energie, Wilhelm-Conrad-R\"{o}ntgen-Campus BESSY II, Albert-Einstein-Str. 15, 12489 Berlin, Germany}

\author{E.~Schierle}
\affiliation{Helmholtz-Zentrum Berlin f\"{u}r Materialien und Energie, Wilhelm-Conrad-R\"{o}ntgen-Campus BESSY II, Albert-Einstein-Str. 15, 12489 Berlin, Germany}

\author{E.~Weschke}
\affiliation{Helmholtz-Zentrum Berlin f\"{u}r Materialien und Energie, Wilhelm-Conrad-R\"{o}ntgen-Campus BESSY II, Albert-Einstein-Str. 15, 12489 Berlin, Germany}

\author{J.-S.~Lee}
\affiliation{Stanford Synchrotron Radiation Lightsource, SLAC National Accelerator Laboratory, Menlo Park, CA 94025, USA}

\author{H.~Jang}
\affiliation{Stanford Synchrotron Radiation Lightsource, SLAC National Accelerator Laboratory, Menlo Park, CA 94025, USA}

\author{H.~Huang}
\affiliation{Stanford Synchrotron Radiation Lightsource, SLAC National Accelerator Laboratory, Menlo Park, CA 94025, USA}

\author{M.~Hashimoto}
\affiliation{Stanford Synchrotron Radiation Lightsource, SLAC National Accelerator Laboratory, Menlo Park, CA 94025, USA}

\author{D.-H. Lu}
\affiliation{Stanford Synchrotron Radiation Lightsource, SLAC National Accelerator Laboratory, Menlo Park, CA 94025, USA}

\author{D.~Song}
\affiliation{National Institute of Advanced Industrial Science and Technology, Tsukuba 305-8565, Japan}

\author{Y.~Yoshida}
\affiliation{National Institute of Advanced Industrial Science and Technology, Tsukuba 305-8565, Japan}

\author{H.~Eisaki}
\affiliation{National Institute of Advanced Industrial Science and Technology, Tsukuba 305-8565, Japan}

\author{Z.-X.~Shen}
\affiliation{Department of Applied Physics, Stanford University, Stanford, California 94305, USA}
\affiliation{Stanford Institute for Materials and Energy Sciences, SLAC National Accelerator Laboratory, 2575 Sand Hill Road, Menlo Park, California 94025, USA}

\author{R. J.~Birgeneau}
\affiliation{Department of Physics, University of California, Berkeley, California 94720, USA}

\author{M.~Yi}
\email[]{mingyi@berkeley.edu}
\affiliation{Department of Physics, University of California, Berkeley, California 94720, USA}

\author{A.~Frano}
\email[]{afrano@ucsd.edu}
\affiliation{Department of Physics, University of California, San Diego, La Jolla, California 95203, USA}

\begin{abstract}
%We have used high-resolution inelastic x-ray scattering (IXS) in combination with Raman scattering to present an experimental survey of the lattice dynamics of bilayer superconducting cuprate compounds. 

We report a persistent low-energy phonon broadening around $q_{B} \sim 0.28$ r.l.u. along the Cu-O bond direction in the high-\textit{T$_c$} cuprate Bi$_2$Sr$_2$CaCu$_2$O$_{8+\delta}$ (Bi-2212). We show that such broadening exists both inside and outside the conventional charge density wave (CDW) phase, via temperature dependent measurements in both underdoped and heavily overdoped samples. Combining inelastic hard x-ray scattering, diffuse scattering, angle-resolved photoemission spectroscopy, and resonant soft x-ray scattering at the Cu \textit{L$_3$}-edge, we exclude the presence of a CDW in the heavily overdoped Bi-2212 similar to that observed in the underdoped systems. Finally, we discuss the origin of such anisotropic low-energy phonon broadening, and its potential precursory role to the CDW phase in the underdoped region.

\end{abstract}

\maketitle

\section{Introduction}

% this is the OD53 formular
%  Bi$_2$Sr$_2$CaCu$_2$O$_{8+\delta}$ 
% and this is the UD32: Bi$_2$Sr$_2$(Ca,Dy)Cu$_2$O$_{8+\delta}$ 

\begin{figure*}[htb]
\captionsetup{justification=raggedright}
\includegraphics[width=2.1\columnwidth]{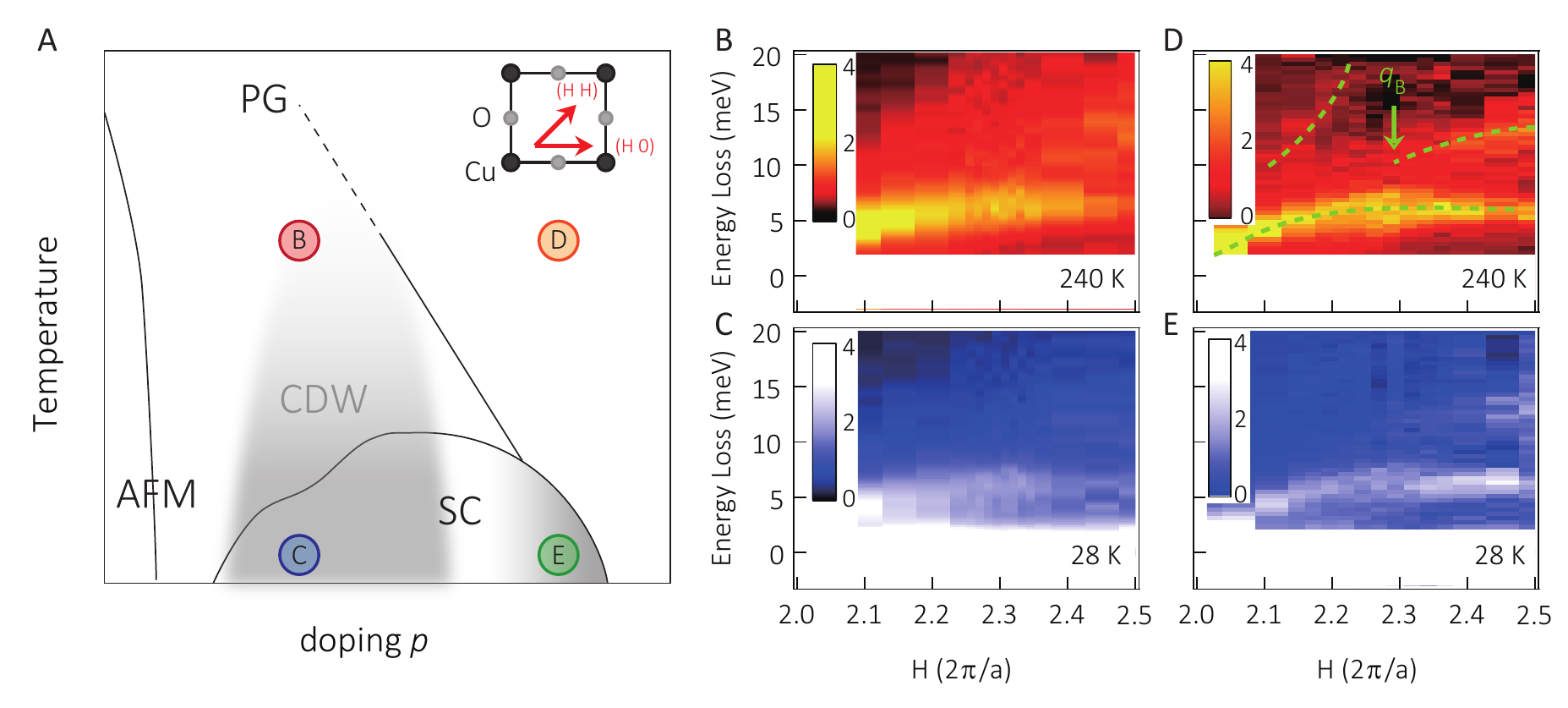}
\caption{(a) Generic phase diagram for the Bi-2212 system, where the grey shade indicates the hitherto-known charge density wave region. The colored circles indicate the four measurement conditions. The inset shows the tetragonal CuO$_2$ unit cell convention used in this study, where \textit{H} is parallel to the Cu-O bond direction. Abbreviations are: AFM - long range antiferromagnetic order, PG - pseudogap, CDW - charge density wave, SC - superconductivity. (b)(c) Low-energy longitudinal phonon dispersion along the (\textit{H} 0 0) direction for conditions B and C in panel (a), where doping is 7\% (UD32). %The intensity is normalized to the signal strength within E$~\in~$[3,20]~meV for visualization. 
(d)(e) Low-energy longitudinal phonon dispersion for conditions D and E in panel (a), where doping is 23\% (OD53). The green dashed lines are guides to the eye, and \textit{q}$_B$ indicates the phonon broadening momentum.}
% highlight the OD region 
% AFM label
\label{fig:figure1}
\end{figure*}

The phase diagram of the high-\textit{T$_c$} cuprate superconductors has many intertwined broken-symmetry states. At the heart of them lies the superconductivity (SC), whose origin has eluded our understanding amid many adjacent phases (Fig.~\ref{fig:figure1}(a))~\cite{keimer2015}. Observations of one-dimensional spin stripe order in La$_{1.48}$Nd$_{0.4}$Sr$_{0.12}$CuO$_4$ (Nd-LSCO) from inelastic neutron scattering (INS) first provided experimental evidence for a competing spin density wave order inside the superconducting phase ~\cite{tranquada1995}. Later, a concomitant pattern of charge stripes with half the period was discovered using resonant x-ray scattering ~\cite{abbamonte2002}. Recent discoveries of charge order in all families of underdoped cuprates have raised a central question ~\cite{shen2005, wu2011, ghiringhelli2012, chang2012, da2013, comin2014} - what is the cause of the charge order?

%On one hand, 
The effect of electronic correlations has been put forward as one leading explanation. The charge order momentum \textit{q}$_{co}$ appears to coincide with a \textit{q}-vector that bridges regions of the partially gapped Fermi surface in the pseudogap state, facilitating an ostensible nesting condition ~\cite{comin2014, shen2005, da2013}. However, this simple nesting picture cannot reconcile with a series of mismatched phononic and electronic properties in Bi-2212 ~\cite{da2013, zhu2015, chaix2017}. Numerical solutions to the 2D Hubbard model show that hole-doping in a half-filled Mott system introduces an inherent electronic instability which breaks translational symmetry even at high temperatures, where the valence electrons crystallize into stripes of various correlation lengths ~\cite{huang2017, zheng2017}.

On the other hand, anomalies in the phonon spectra have been widely observed around \textit{q}$_{co}$, suggesting an intimate role of the lattice in the charge ordering phenomenon. In underdoped YBa$_2$Cu$_3$O$_{6+y}$ (YBCO), the low-energy phonon dispersions show anomalous softening and broadening behavior at \textit{q}$_{co}$, both of which respond strongly to the superconducting T$_c$~\cite{letacon2014}. It's also proposed that this softening and broadening may also arise from a combined effect of multiple phonon branches ~\cite{blackburn2013}. Single layer compound La$_{2-x}$Ba$_{x}$CuO$_4$ also shows low-energy phonon broadening near \textit{q}$_{co}$ at $x=0.125$~\cite{miao2018}. In addition to low-energy phonons, a recent resonant inelastic x-ray scattering (RIXS) study of underdoped Bi-2212 shows that a $\sim$60~meV oxygen breathing phonon also significantly softens and broadens near \textit{q}$_{co}$ up to 240~K ~\cite{chaix2017}.

% Phonon anomalies are also important in the context of SC. 
The relation of lattice vibrations to superconductivity is also highly relevant. Low-energy electron dispersion anomalies have been observed in Bi$_2$Sr$_2$CuO$_{6+\delta}$ (Bi-2201), Bi-2212 and La$_{2-x}$Sr$_{x}$CuO$_4$ (LSCO) via angle-resolved photoemission spectroscopy (ARPES) ~\cite{vishik2010, anzai2010, kondo2013, anzai2017}, where a full hierarchy of bosonic modes imprinting on the electronic structure ~\cite{lanzara2001, cuk2004, dahm2009}. Among them, two low-energy phonon modes are proposed to have the ability to enhance the \textit{d}-wave Cooper pairing, thus boosting the superconducting T$_c$ ~\cite{johnston2010, johnston2012}. The notion is further bolstered by recent reports of enhanced T$_c$ induced by strong electron-phonon coupling (EPC) at the FeSe-SrTiO$_3$ interface~\cite{lee2014, li2016, song2017}. 

Despite the importance of knowing details of the phonon spectrum of Bi-based cuprate superconductors, there are only limited reports. Phonon measurements by inelastic neutron scattering have mainly focused on systems where large bulk crystals are available, including YBCO ~\cite{baron2008, raichle2011} and Hg-1201 ~\cite{dastuto2003, uchiyama2004}, both of which lack systematic comparisons to ARPES and scanning tunneling microscopy (STM). Recent advances in high-resolution inelastic x-ray scattering (IXS) enables investigations of materials with limited sample volume such as crystals of the single layer compound Bi-2201 and its bi-layer counterpart Bi-2212 ~\cite{graf2008, bonnoit2013}. A recent INS study reports putative low-energy phonon anomalies around \textit{q}$_{co}$ in optimally doped Bi-2212 ~\cite{merritt2017}, but leaves open the discussion about their origin. Given this observation, we raise the following questions: is there a connection between this anomaly and the CDW observed in underdoped Bi-2212? More fundamentally, is the phonon anomaly playing a causal role in the formation of CDW and the enhancement of superconductivity in the high-\textit{T$_c$} cuprates?

To gain insights into these open questions, we hereby report a comparative series of photon-scattering experiments performed on samples of underdoped (hole doping $p$ = 0.07) and heavily overdoped ($p$ = 0.23) Bi-2212. We utilize high-resolution IXS to measure the low-energy phonon excitation spectrum, and find a persistent broadening near the \textit{q}$_{co}$ for both dopings. To inspect whether there is CDW in the overdoped regime, we conducted a thorough set of diffraction experiments surveying the reciprocal space of the overdoped sample. These include diffuse scattering (DS), IXS, and (energy integrated) resonant elastic x-ray scattering (REXS). We observe no diffraction signal of CDW analogous to that of overdoped Bi-2201 ~\cite{peng2017} or any other underdoped cuprates. Importantly, there is a region of reciprocal space where the longitudinal acoustic (LA) phonon branch crosses an extremely low-energy optical branch, potentially creating a broad hybridization zone that encompasses the charge ordering \textit{q}$_{co}$. We speculate that the low-energy phonon broadening may act as a lattice precursor, which, when coinciding with strong electronic correlation effects in the underdoped cuprates (e.g. the pseudogap), can result in CDW.

This paper is organized as follows. First, we discuss the details of the various experimental techniques and sample information. Then, we present the IXS data from both samples and temperatures. We discuss our analysis of the phonon excitation spectrum, the fitting model used, and the results. Then, we argue for a null observation of a CDW order in the overdoped Bi-2212 based on REXS, IXS and DS measurements. A discussion and outlook is lastly presented, highlighting the implications of our findings as well as an outlook of how our hypothesis could be further tested.

\section{METHODS}

Single crystals of Bi-2212 were grown at AIST, Japan with the optical floating zone method. Underdoping was achieved via partial replacement of Ca with Dy to form Bi$_2$Sr$_2$(Ca,Dy)Cu$_2$O$_{8+\delta}$. Heavy overdoping was achieved by annealing the sample in a 400~bar, high pressure oxygen environment to form super-oxygenated Bi$_2$Sr$_2$CaCu$_2$O$_{8+\delta}$. The superconducting transition temperatures were then confirmed using a Quantum Design SQUID magnetometer. Throughout this paper, we refer to the underdoped sample (T$_c$=32~K) as UD32, and to the overdoped sample (T$_c$=53~K) as OD53. The lattice constants at 300~K, determined using the multiple scattering techniques described below, are $a\approx b$=3.829~\AA, $c$=30.45~\AA~ for UD32, and $a\approx b$ 3.825~\AA, $c$=30.81~\AA~ for OD53 in a tetragonal cell convention (Fig.~\ref{fig:figure1}(a)). The reciprocal lattice units throughout this paper are based on this convention (\textit{q}$_x$=$\frac{2\pi}{a}$*\textit{H}, \textit{q}$_y$=$\frac{2\pi}{b}$*\textit{K}, \textit{q}$_z$=$\frac{2\pi}{c}$*\textit{L}).

To characterize the phonon spectra, we employed IXS and Raman scattering spectroscopy. The IXS measurements on UD32 were done at the ID28 beamline of the European Synchrotron Radiation Facility (ESRF) with 17.79~keV (instrument energy resolution $\Delta$E$\sim$3.0~meV) and 23.73~keV ($\Delta$E$\sim$1.5~meV) incident photons. The IXS measurements on OD53 were done at both ESRF and the 30-ID HERIX beamline at the Advanced Photon Source, Argonne National Laboratory, where the photon energy used was 23.724~keV with an energy resolution of $\Delta$E$\sim$1.5~meV. In both cases the samples were glued to a sample holder inside a closed-cycle cryostat on a 4-circle goniometer. The IXS experiments were all performed in constant wave-vector mode using transmission geometry. Every inelastic spectrum scan was performed to include the elastic line as a calibration reference for energy drift corrections ~\cite{krisch2006, baron2015}. Room-temperature mixed-polarization Raman scattering experiments were carried out on both samples at the Stanford Nano Shared Facility at Stanford University using a Horiba Labram Raman spectrometer with 1.96~eV incident photons. The experiment geometry, in Porto notation, was $\bar{z}(xu)z$ and $\bar{z}(yu)z$, where u means no polarization analysis for the outgoing light. Diffuse scattering measurements were done at the ID29 beamline at the ESRF with 12.65~keV photons, where the temperature control was achieved with a nitrogen cryostream down to 120~K. The resonant soft x-ray scattering experiments were performed at the UE46-PGM1 beamline of the Helmholtz Zentrum Berlin, Germany and at beamline 13-3 of the Stanford Synchrotron Radiation Lightsource (SSRL). In both cases, the photons were resonant to the Cu \textit{L$_3$}-edge (931.5~eV), and the scattering geometries were identical to previous studies of charge ordering in cuprates ~\cite{frano2014technique}. Angle-resolved photoemission (ARPES) measurements were done at SSRL beamline 5-4 using 18.4~eV incident photon energy with the linear photon polarization aligned with the Cu-O bond direction. The doping of each sample was determined using the parabolic \textit{T$_c$}-doping relation ~\cite{presland1991}.

\begin{figure*}[htb]
 	\captionsetup{justification=raggedright,width=2.1\columnwidth}
 	\includegraphics[width=2.1\columnwidth]{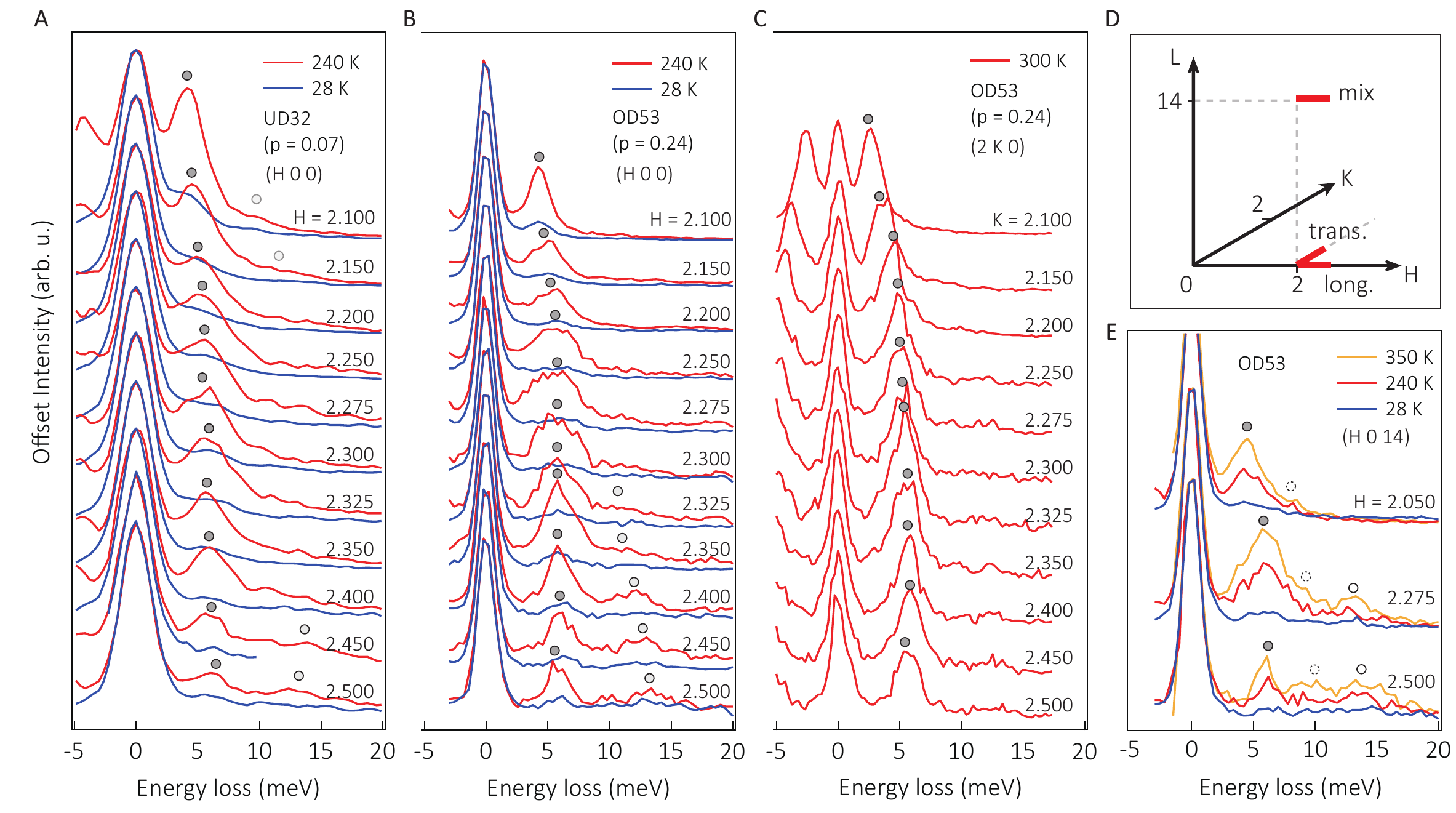}
 	\caption{Normalized IXS energy-loss scans. Longitudinal cuts along (\textit{H} 0 0) at 240~K (red) and 28~K (blue) for (a) UD32 ($\Delta$E$\sim$3.0~meV) and (b) OD53 ($\Delta$E$\sim$1.5~meV). (c) Transverse cuts along (2 \textit{K} 0) at 300~K in OD53. (d) Schematics of the scan directions in reciprocal space. (e) Line cuts for OD53 with a mixture of transverse and longitudinal channels along (\textit{H} 0 14). All spectra displayed here are normalized to the elastic peak intensity and offset for clarity.}
 	\label{fig:figure2}
 	% remove error for Raman
 	% arrow to the Raman peaks
 	% limit elaboration on the Raman peaks ~ 14meV
 	% Raman label keeper?
 	% check color of the labels, recolor the Raman after error bar removal
 	% guide to the eye, grey keeper?
 	% offset intensity -> Raman intensity
 	% Raman IXS IXS
 	
\end{figure*}

\section{Results}
\subsection{Phonon spectra}
Figure \ref{fig:figure1}(a) displays a generic phase diagram of the superconducting cuprates with four points labeled using letters B-E, corresponding to the doping/temperature conditions for which the phonon dispersion spectra were studied. Two samples, UD32 and OD53, were mainly probed at two different temperatures: 28~K and 240~K. Only the low-temperature measurement of UD32 falls within the CDW region (point C). The OD53 measurements (points D and E) were chosen so to be far from the pseudogap phase. Figure~\ref{fig:figure1} (b-e) correspondingly display intensity color maps of the low-energy phonon spectra stitched together from individual IXS energy-loss cuts at different momenta along the longitudinal (\textit{H} 0 0) direction for each of these conditions. The window of energy $\Delta$E$~\in~[-1,2]$~meV contains the elastic contribution of the spectrum and has been masked out for clarity. A strong elastic Bragg peak lies at (2 0 0) which floods the inelastic spectrum, precluding the measurements near the zone center.

Let us now discuss the phonon branches away from the zone center discernible in all spectra. First, a streak of intensity within the energy range $E\in~[3,7]$~meV can be seen in all four panels. This low-energy phonon mode has been observed by INS in Bi-2212 and by IXS in YBCO ~\cite{merritt2017,blackburn2013,letacon2014}. Second, a branch with energy $E\in~[10,15]$~meV can be resolved in all four spectra for \textit{H}$~\in~[2.2,2.5]$~r.l.u. Finally, a weak signal can be identified at high energies $E\gtrsim$~10~meV for \textit{H}~$\in~[2.1,2.25]$~r.l.u., matching an \textit{A}$_{1g}$ phonon found in calculations ~\cite{uchiyama2004}. At 240~K (Figure \ref{fig:figure1}(b,d)), the intensity of those phonon branches are more prominent due to the Bose factor. We point out a broadening of the linewidth of the low-energy mode near $q_B\in[2.2,2.35]$~r.l.u., particularly obvious for the high temperature spectra in OD53, which requires a closer look. %can we trace out the dispersion in the color plot? the weaker feature is hard to see

% The broadening in  Fig 1 B/D is not very clear if the readers don't read the text and see the detailed cuts in Fig 2. We could either put vertical bar indicating the width along the dispersion or trace out the boundary of the width to show the broadening effect. 

% Another option is to put recent HH0 color plot by side. From that the broadening may be more clear by contrast with non-broadening along the diagonal direction. 
% More thought for the second option: If so, it is perhaps too crowded in Fig 1. The HH0 contour plot can be put separately in the discussion section later with detailed cuts along HH0. 

\begin{figure*}[htb]
 	\captionsetup{justification=raggedright,width=2.0\columnwidth}
 	\includegraphics[width=2.0\columnwidth]{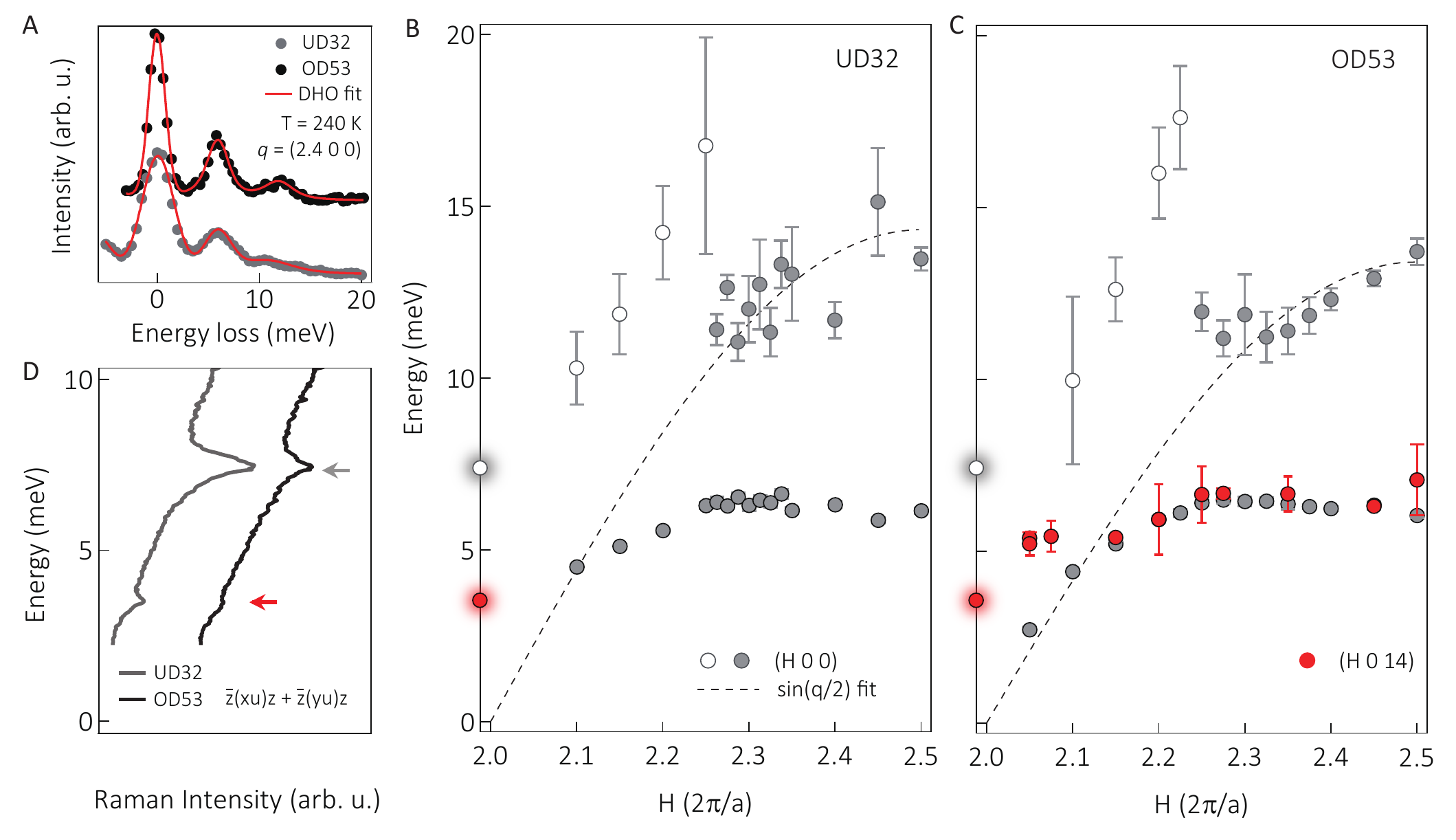}
 	\caption{Fitted dispersion. (a) Examples of damped harmonic oscillator fitting for two samples, convolved with their respective instrument resolution. Fitted peak positions for IXS data on (b) UD32 (c) OD53 Bi-2212. All IXS data shown here are taken at 240~K except for the (H 0 14) scans at 300~K. Dashed lines in (b) and (c) are fit to the acoustic mode dispersion with sin($qa$/2) assuming the higher energy mode at the high $q$ being from the longitudinal acoustic branch. The markers at \textit{q}$ = (2~0~0)$ denote energies at 3.5~meV and 7.5~meV derived from room tempearture Raman measurements on UD32 (grey) and OD53 (black) as shown in (d). Note the significant presence of the optical mode evident in the red IXS data point at (2.05 0 14), which is very close to the zone center.}
 	\label{fig:figure3}
\end{figure*}

% It is important to highlight  which LA branch crosses a low-energy optical phonon in Fig 3.C. Since it is the position potentially produce the broadening at q_c as you claimed in the summary at the beginning. 

%Discuss in regions in reciprocal space where the cuts are taken (DS map), and highlight the main point of this paper: broadening.
In Figure~\ref{fig:figure2}, we present raw individual energy-loss spectra at varying reciprocal space directions shown schematically in Fig.~\ref{fig:figure2}(d) by solid red bars. Purely longitudinal cuts were taken along the \textit{H}-direction near (2+\textit{h} 0 0), while a cut with mixed longitudinal and transverse components was taken along the \textit{H}-direction near (2+\textit{h} 0 14). Purely transverse cuts were taken along the \textit{K}-direction near (2 \textit{K} 0) direction but only for OD53. Panels (a) and (b) show the raw inelastic data of the longitudinal cuts for UD32 and OD53, with 3~meV and 1.5~meV energy resolution respectively. The energy-loss spectra are shifted vertically for clarity. High (red) and low (blue) temperature data are compared for each sample.
% panel (d) and (e) may need to include recent data. 
For OD53, panel (c) shows the transverse cut measured at T=300~K, and panel (e) shows the high \textit{L} cut taken at three different temperatures at the labeled \textit{H}-positions. The circles mark the positions of possible phonon modes. And the broadening of the low-energy longitudinal branch (Fig.~\ref{fig:figure2}(a) and (b)) readily contrasts the lack of that in the transverse branch (Fig.~\ref{fig:figure2}(c)). A similar broadening has been observed in the low-energy acoustic mode of YBCO for temperatures below the onset of the CDW, and maximizes around the superconducting T$_c$ ~\cite{letacon2014,blackburn2013}. However, the broadening effect we hereby report can be observed in both underdoped and overdoped samples, for both high and low temperatures (Fig.~\ref{fig:figure1}(a)). Particularly interesting is the fact that the effect is clearly manifested in OD53 - a phase region far away from any common electronic order instability or anomalous Fermi surface phenomena, for example, the pseudogap. We note a recent INS report also observed such broadening for optimal doping at room and low temperatures in Bi-2212 ~\cite{merritt2017}.% This may be put in the discussion section to avoid attracting the attention from the other work. 

%To obtain a more qualitative view of this, we fit the modes using a damped SHO with Bose-Einstein occupation correction. Fig 3: fit of energy of all modes also taking T+L mix. 
\begin{figure}[htb]
 	\captionsetup{justification=raggedright,width=1\columnwidth}
 	\includegraphics[width=1\columnwidth]{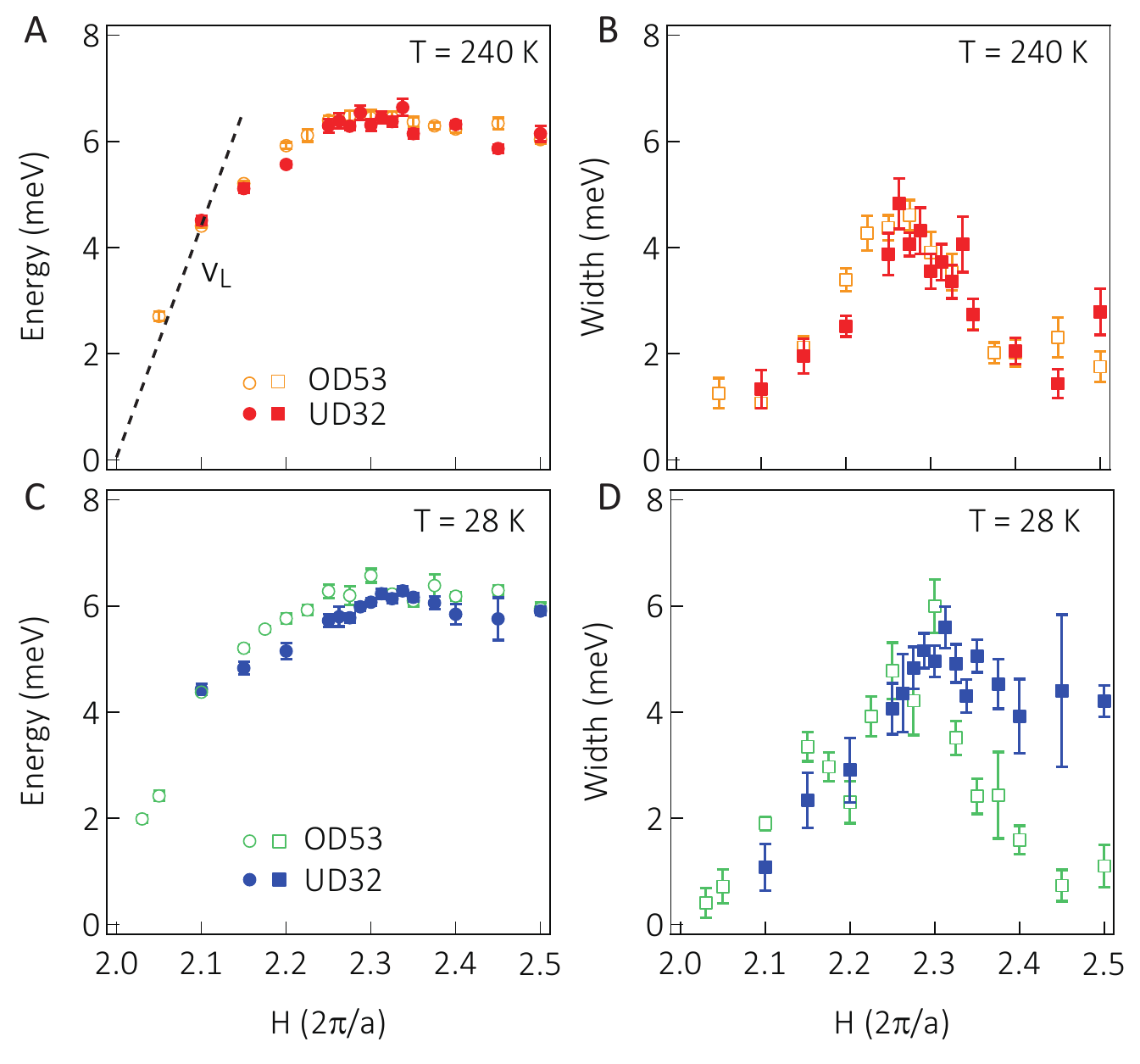}
 	\caption{(a) High temperature dispersion for UD32 and OD53 along (\textit{H}~0~0) direction. (b) High temperature phonon width. (c) Low temperature dispersion for UD32 and OD53. (d) Low temperature phonon width. Dispersion is fitted with damped harmonic oscillator model, with the Bose-factor considered. The black dashed line references the bulk in-plane longitudinal sound velocity at 4370~m/s.}
 	\label{fig:figure4}
\end{figure}

\begin{figure}[htb]
 	\captionsetup{justification=raggedright,width=1\columnwidth}
 	\includegraphics[width=1\columnwidth]{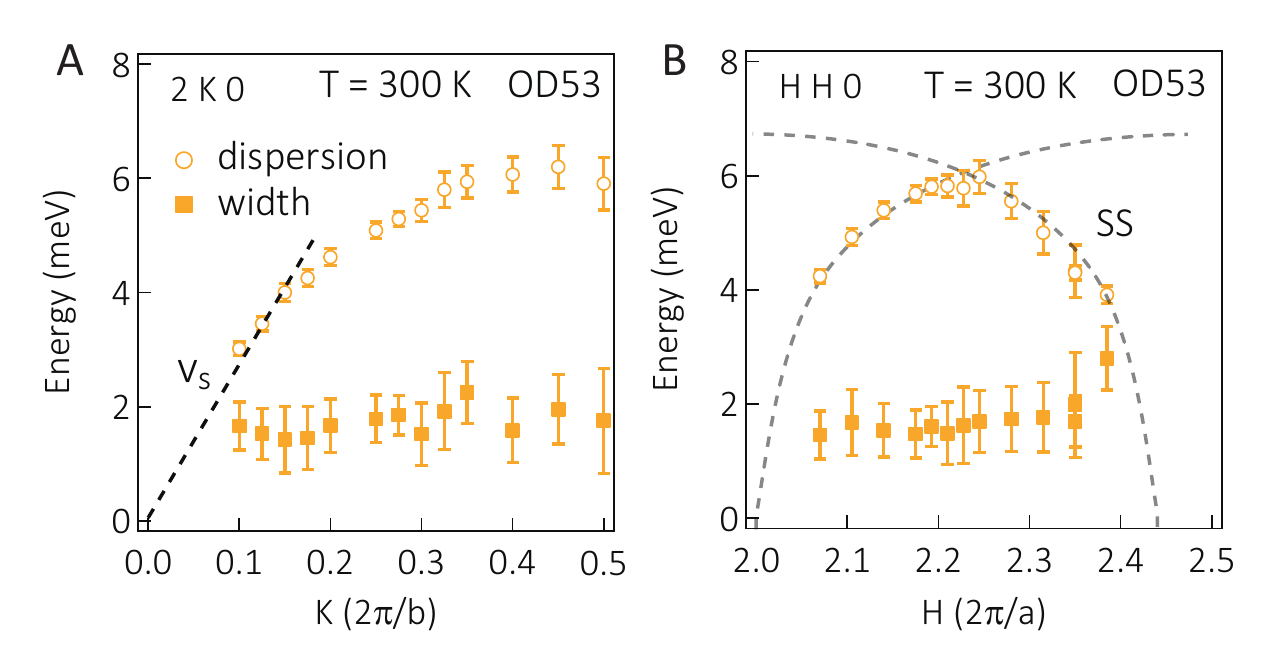}
 	\caption{(a) Transverse acoustic branch dispersion (circle) and its width (square) for OD53 at 300~K. The black dashed line references the bulk in-plane fast shear sound velocity at 2460~m/s. (b) Longitudinal diagonal (Cu-Cu) dispersion and its width for OD53 at 300~K. The grey dashed line indicates the satellite branch originated from the incommensurate superstructure (SS) Bragg peak.}
 	\label{fig:figure5}
\end{figure}

% Figure 4 needs to have better arrangement. Different temperature, different doping and different Q are confusing to put together. It may be better to have separate figures. 

%Highlight the observation of a mode near zone center. To corroborate such a branch, we do Raman scattering on both samples and clearly see a mode near 3.5meV on both samples.

% This paragraph is lack of well organizing. Probably we could separate it into two paragraphs. Firstly, to show the match of low-energy optical mode with Ramen scattering.   Secondly, to show the qualitative analysis of the broadening and how tconvolvewo branches (one LA and low-energy optical mode) crosses at q_c. 

To obtain quantitative information about the broadening of the low-energy phonon, the energy-loss spectra were fitted to the resolution-convolved damped harmonic oscillator functions with the Bose factor correction~\cite{budai2014}. Fig.~\ref{fig:figure3}(a) shows two typical fits with a single Gaussian and two Bose-factor corrected damped harmonic oscillator functions, all convolved with the respective energy resolution. Figure~\ref{fig:figure3} (b) and (c) present the dispersions for the low-energy modes. The IXS spectral fitting results in the longitudinal geometry for UD32 and OD53 samples are shown in panels (b) and (c), respectively. Panel (c), in addition, displays the fitted energy positions of the low-energy modes from the mixed geometry (\textit{H} 0 14) as red circles. Since the zone-center IXS measurements are flooded by the Bragg intensity at (2 0 0), Raman spectroscopy was used to determine the mode energies at \textit{q}$\sim 0$. The room-temperature Raman spectra for both samples are shown in panel (d). The dispersive nature of three modes can thus be identified.
The low-energy \textit{A}$_{1g}$ optical mode starts for both samples at E=7.5~meV at the zone center (Fig.~\ref{fig:figure3}(d)), and quickly trends upwards with increasing in-plane momentum (Fig.~\ref{fig:figure3}(b,c), open circles). The $\sim$ 6~meV zone boundary phonon may initially appear as a natural candidate for the acoustic branch, as implied in an early Hg-1201 study ~\cite{uchiyama2004}. Meanwhile, calculations in LSCO, NCCO and Bi-2212 all show acoustic branches extending up to $\sim$13-16~meV at the zone boundary ~\cite{chaplot1995,d2002,shimada1998}. A simple diatomic spring-ball model yields a linear to sinusoidal small-\textit{q} dispersion relations for acoustic modes, whose zero-\textit{q} slope can be cross-checked with the bulk sound velocity. The dashed lines shown in panels (b) and (c) are the result of fitting the higher energy large-q mode (for \textit{H}$~\in~[2.2,2.5]$~r.l.u.) to a sinusoidal dispersion relation sin($qa$/2), with the energy at the zone boundary around 13-14~meV. This LA branch intersects the low-energy branch at \textit{H}$\sim$2.1~r.l.u., with a corresponding longitudinal sound velocity of 4360$\pm$390~m/s, consistent with both inelastic neutron scattering result in optimally doped Bi-2212 single crystals ~\cite{merritt2017}, as well as the bulk ultrasound measurement of the in-plane longitudinal sound velocity of 4370$\pm$10~m/s in very dense textured Bi-2212 ~\cite{chang1993}. On the other hand, a simple sinusoidal dispersion model does not appropriately describe the low-energy mode. Instead, the low-energy mode appears to intersect the zone center at finite energy. The mixed polarization geometry allows the low-energy mode to be clearly seen at non-zero energy close to the zone center at \textit{H}=2.05~r.l.u. (red circles in panel (c)). Furthermore, the Raman data of panel (d) shows an optical excitation at 3.5~meV which extrapolates well onto this branch. Thus, this flat low-energy optical mode may participate in the temperature- and doping-independent broadening effect that we observe.

%Discuss now the width of the acoustic modes: broadening! Figure 4. Compare the two dopings side-by-side: central message of this paper. No temperature dependence.

% This paragraph could be put in the discussion section. 
We present a quantitative analysis of this branch as a function of sample temperature in Figure~\ref{fig:figure4}. The fitted energy (panels (a) and (c)) and mode linewidth (panels (b) and (d)) are plotted for both samples at high and low temperature, as labeled. The linewidths increase from $\sim$2~meV near the zone center to $\sim$6~meV for \textit{H}$~\in~$[2.2,2.35]~r.l.u., and has no appreciable temperature dependence. Similarly, the energy of the mode shows no change as a function of temperature. Previous reports on YBCO revealed a broadening of a low-energy acoustic phonon at the onset of CDW order for temperatures below T$\sim$120~K ~\cite{blackburn2013,letacon2014}. No softening is observed at low temperature for UD32, where the CDW has been observed using soft x-ray resonant scattering at the Cu \textit{L$_3$}-edge ~\cite{da2013}. However, the charge modulations observed by REXS in Bi-2212 are comparatively weaker and of shorter range than those in YBCO. The response of the lattice vibration near \textit{q}$_{co}$ to such a subtle CDW effect might also be weaker in Bi-2212.

The broadening also appears to be a unique property for the longitudinal branch along the Cu-O bond direction in Bi-2212. The transverse branch (Fig.~\ref{fig:figure5}(a)) and the longitudinal, diagonal dispersion (perpendicular to the superstructure direction, Fig.~\ref{fig:figure5}(b)) do not show any appreciable momentum-dependent broadening. This also helps rule out the trivial superstructure-induced phonon broadening, in which case a broadening would have occurred where the superstructure phonons intercept. It should also be noted that the in-plane transverse branch also yields a sound velocity of 2560$\pm$150~m/s, in good agreement with the bulk in-plane fast shear sound velocity of 2460$\pm$10~m/s measured with ultrasound ~\cite{chang1993}.

\subsection{CDW search}

\begin{figure*} 
 	\captionsetup{justification=raggedright,width=2.0\columnwidth}
 	\includegraphics[width=2.0\columnwidth]{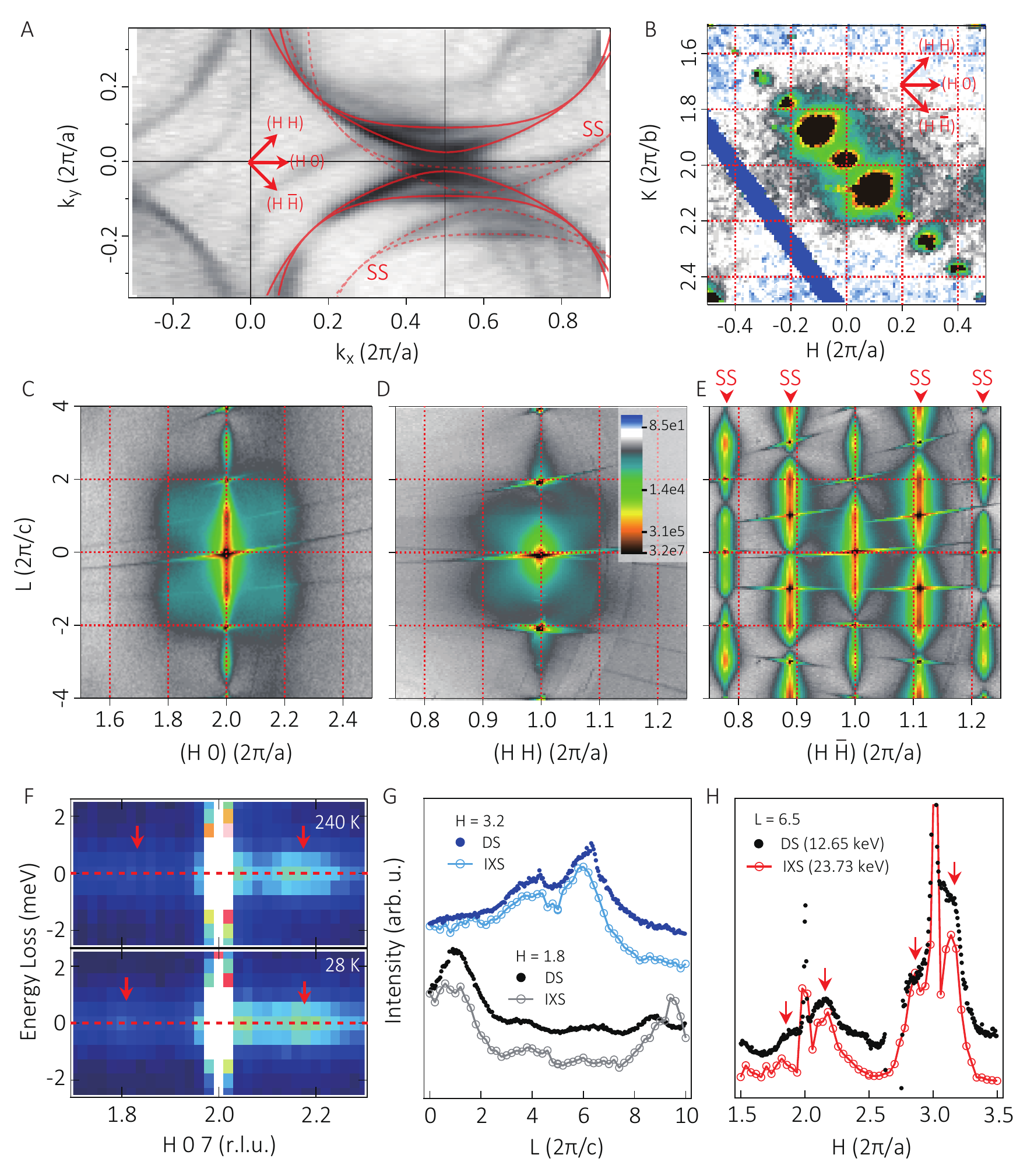}
 	\caption{Looking for evidence of CDW in OD53. (a) ARPES Fermi surface map at 104K. The solid and dashed red lines indicate the Fermi surface from the main bands and the supersturcture bands, respectively. (b) DS at (\textit{H} \textit{K} 1) plane, where a significant intensity tail from the Bi-O superstructure (SS) can be seen even along (\textit{H} 0) direction. (c) 'butterfly' pattern in (\textit{H} 0 \textit{L}) plane. (d) Similar pattern perpendicular to the SS in (\textit{H} \textit{H} \textit{L}) plane). (e) DS reciprocal space map along the SS direction shows that the intensity of the first order SS Bragg peak maximizes at \textit{L} = 1. (f) IXS spectra of the satellite peaks along (\textit{H} 0) show their highly elastic nature. (g)(h) A comparison between IXS (zero energy-loss line) and DS (energy integration) shows agreement.}
 	\label{fig:figure6}
\end{figure*}

Given the observation of a lattice vibration anomaly in overdoped Bi-2212 near \textit{q}$_{co}$, we now address in detail our search for possible CDW order in overdoped cuprates. We carried out a thorough survey of the reciprocal space which is shown in Fig.~\ref{fig:figure6}. First of all, Fig.~\ref{fig:figure6}(a) shows an in-plane Fermi surface mapped at 104~K. Full hole-like Fermi surfaces can be clearly seen. This is in agreement with previous reports of a closed Fermi surface in overdoped cuprates ~\cite{YuHe2018_1}. Replicas of these surfaces can be observed along the (\textit{H}~-\textit{H}) direction. These arise from the buckling wave of the Bi-O layer distortion. This superstructure results in satellite peaks with incommensurate vectors splitting from a Bragg peak along the (\textit{H}~-\textit{H}) direction. Fig.~\ref{fig:figure6}(b) shows a diffraction map near the (0 2 1) Bragg peak, where multiple harmonics of these satellites can be observed. Importantly, we note that these superstructure peaks can be broad enough to have the tail of the first order satellite (near the zone center) intersect the (\textit{H} 0) or (\textit{K} 0) lines at reduced in-plane values $\sim$0.18~r.l.u. Panel (c) shows a cut of the (\textit{H} 0 \textit{L}) plane. Sharp, intense Bragg peaks occur at (2 0 \textit{L}=even) and elongated out-of-plane rods can be seen at (2 0 \textit{L}=odd). We focus our attention on the diffuse halos of intensity observed near (2$\pm$0.18 0 $\pm$1), which are reminiscent of the CDW pattern as observed by DS in underdoped YBCO ~\cite{letacon2014}. A similar pattern of diffuse intensity can be observed in panel (d), which is a cut in the (\textit{H} \textit{H} \textit{L}) plane near \textit{H} = \textit{K} = 1 that runs perpendicular to the superstructure. For comparison, panel (e) shows a cut within a plane parallel to the superstructure, i.e. (\textit{H} -\textit{H} \textit{L}), where many of these satellite peaks can be observed, and the \textit{L} = 1 first order superstructure peak intensity overwhelms that of a primary Bragg peak. 
The diffuse intensity observed in panel (c) at (2$\pm$0.18 0 $\pm$1) can also be observed in the elastic part of the IXS scans, as plotted in panel (f). The intensity of the elastic signal increases at \textit{H}=2.18~r.l.u., and persists till 240~K at least. 
Furthermore, the \textit{L}-dependence of these diffuse features is shown in panel (g), which also compares the data measured directly by IXS at zero energy loss to the equivalent cut extracted from the DS maps. Both measurements display very broad, rod-like features. In panel (h) the \textit{H}-dependence of these features can be observed for an intermediate \textit{L}-value, again displaying the elastic IXS data and cuts from the DS map. There is a consistent feature at (\textit{H}$\pm$0.18 0 \textit{L}) around every Bragg position. We measured these diffraction features up to room temperature and did not see any abrupt change in intensity. We speculate, therefore, that the features originate as intensity leakage from the broad Bi-2212 superstructure satellite peaks - which is strongly suggested in panel (b). These features are therefore unlikely CDW peaks of the same nature as observed in the underdoped systems.

\begin{figure*} 
 	\captionsetup{justification=raggedright,width=2.0\columnwidth}
 	\includegraphics[width=2.0\columnwidth]{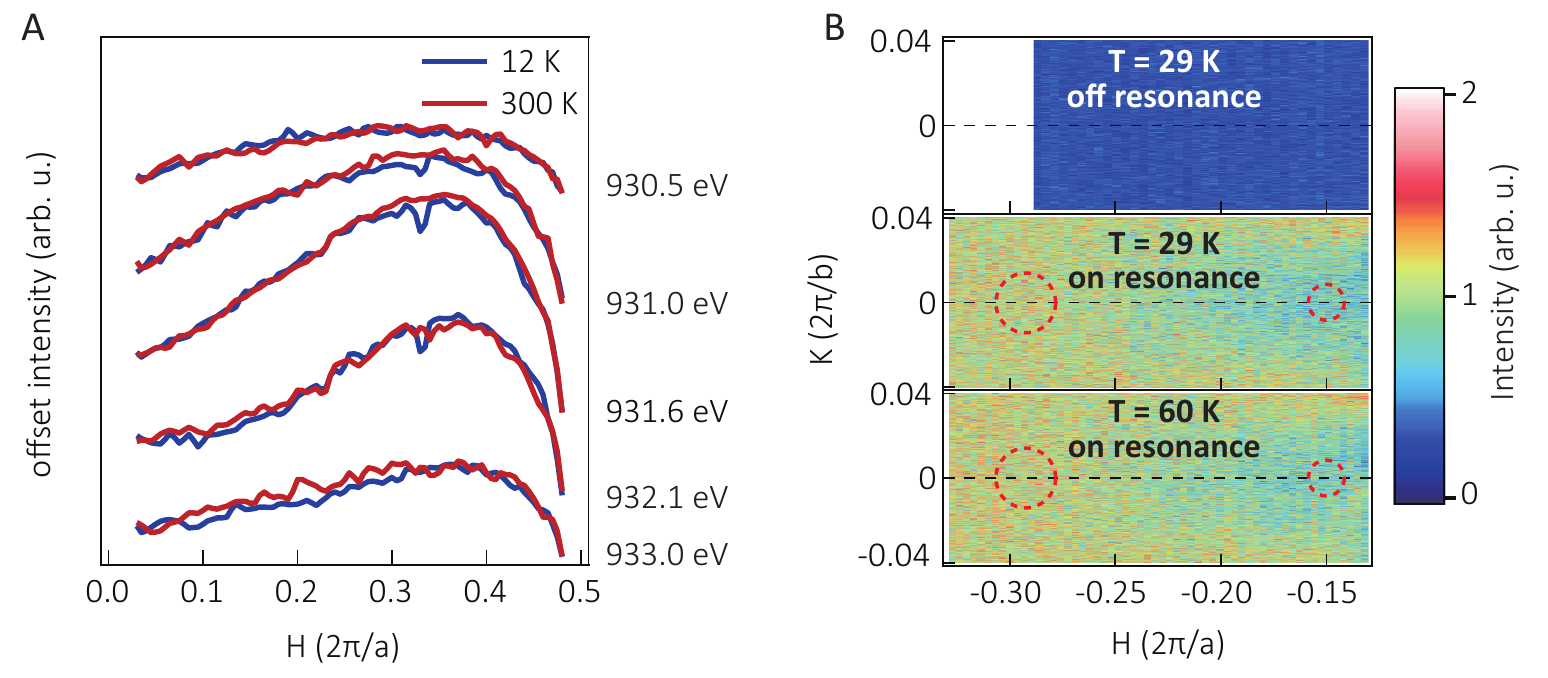}
 	\caption{Lack of resonance and temperature response from the expected CDW-ordering \textit{q} for OD53 system. (a) High temperature (300~K, red) and low temperature (12~K, blue) photon energy dependence near the Cu \textit{L$_3$}-edge via REXS. (b) 2D reciprocal space map in the (\textit{H} \textit{K} $\sim$3.5) plane at the Cu-\textit{L$_3$}-edge also shows a flat background, for both off and on-resonance (top and middle), below and above $T_c$ (middle and bottom). The red dashed circles are where potential CDW signals are anticipated if they were to resemble the underdoped Bi-2212 or overdoped Bi-2201.}
 	\label{fig:figure7}
 	% reference curve for a positive result (UD Bi2212)
\end{figure*}

To further test for the potential presence of weak CDW correlations with very short correlation length in the overdoped sample OD53, we have also performed a series of REXS experiments following a similar procedure as with underdoped Bi2212 and other cuprate families ~\cite{frano2014technique}. Figure~\ref{fig:figure7} shows the results from both a photodiode point detector and a CCD 2D detector. Panel (a) displays rocking curve scans projected onto the \textit{H}-axis, measured with a point detector scattering photons with energies close to the Cu-\textit{L$_3$} absorption edge (E=931.5~eV). The range in \textit{L} of these arched rocking curves is \textit{L}$~\in~$[4.5,1.6]~r.l.u. To identify a potentially weak and short-ranged CDW signal, we compare high and low temperature data. Unlike in the underdoped Bi-based cuprates ~\cite{comin2014, da2013} where at low temperatures the signal shows an increase in intensity near \textit{H}$~\in~$[0.25,0.3]~r.l.u., no intensity excess can be observed in these low-temperature scans for the overdoped system. Figure~\ref{fig:figure7}(b) shows reciprocal space maps of the (\textit{H} \textit{K} 3.5) plane taken with the two-dimensional CCD detector at high/low temperature and on/off resonance. No sign of a peak associated with CDW correlations can be discerned. We have thus surveyed the entirety of reciprocal space afforded by the small Ewald sphere and geometry limitations at Cu-\textit{L$_3$} edge. In summary, using several scattering techniques that reliably detect CDW order in underdoped cuprates, we conclude a null observation of CDW order in overdoped Bi-2212.

\section{Discussion and Outlook}

% In this section, we could put more figures on recent experiment along HH0 to support the second scenario. 

\begin{figure} 
 	\captionsetup{justification=raggedright,width=1.0\columnwidth}
 	\includegraphics[width=1.0\columnwidth]{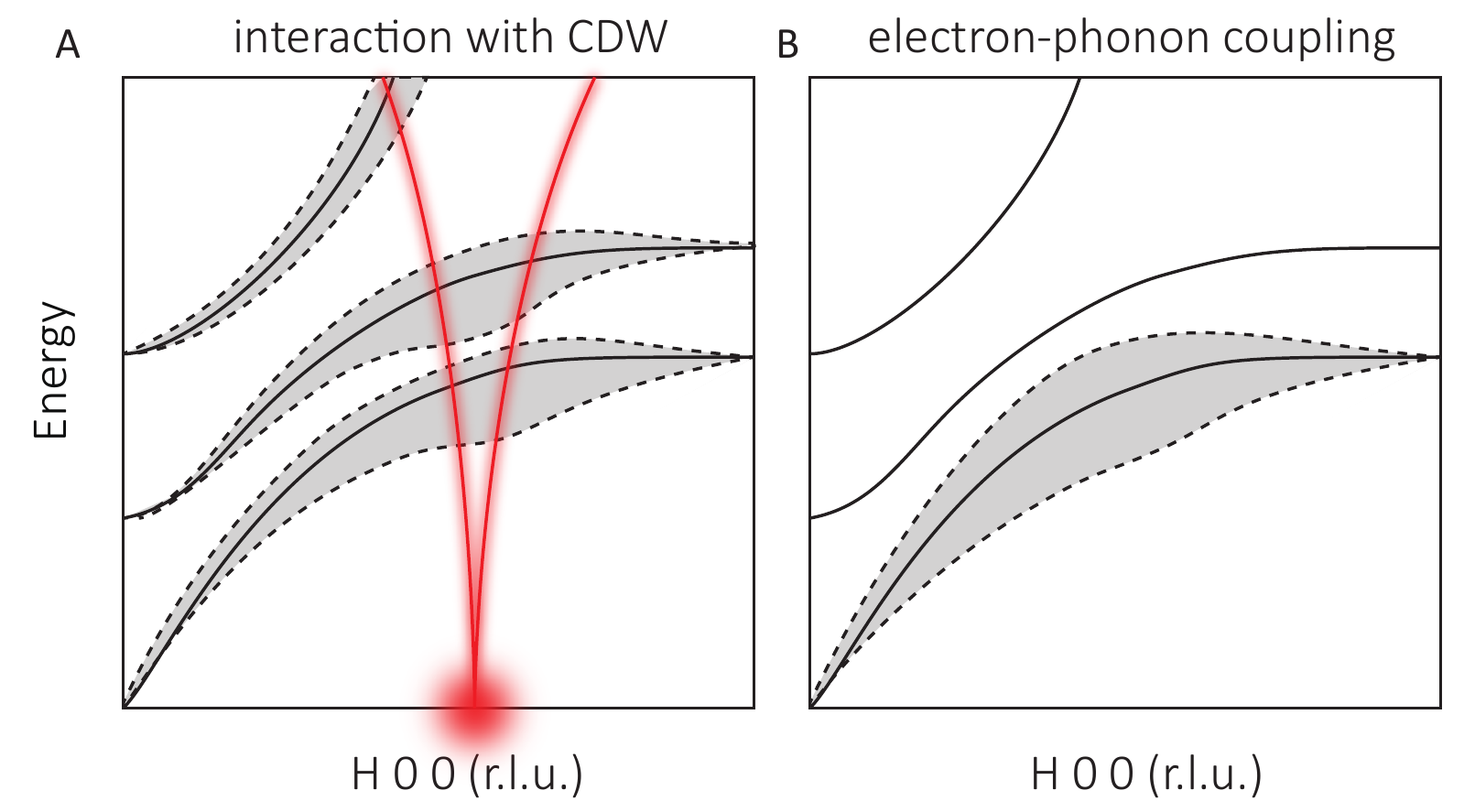}
 	\caption{Two potential scenarios for the broadening (schematics): (a) low-energy phonon interacts with the CDW soft mode.(b) anisotropic electron-phonon interaction due to the possible forward scattering nature of this phonon. The maximum phonon broadening \textit{q}$_B$ roughly agrees with the reported CDW wavevector \textit{q}$_{co}$.}
 	\label{fig:figureScenarios}
 	% nonmonotomically dependent - redundancy
 	% take out colloquial statements
\end{figure}
%does Fig 7 reflect phonons correctly? The low-E optical is below the acoustic mode at larger q, which is not the case in (a) and (c)

\begin{figure} 
 	\captionsetup{justification=raggedright,width=1.0\columnwidth}
 	\includegraphics[width=0.9\columnwidth]{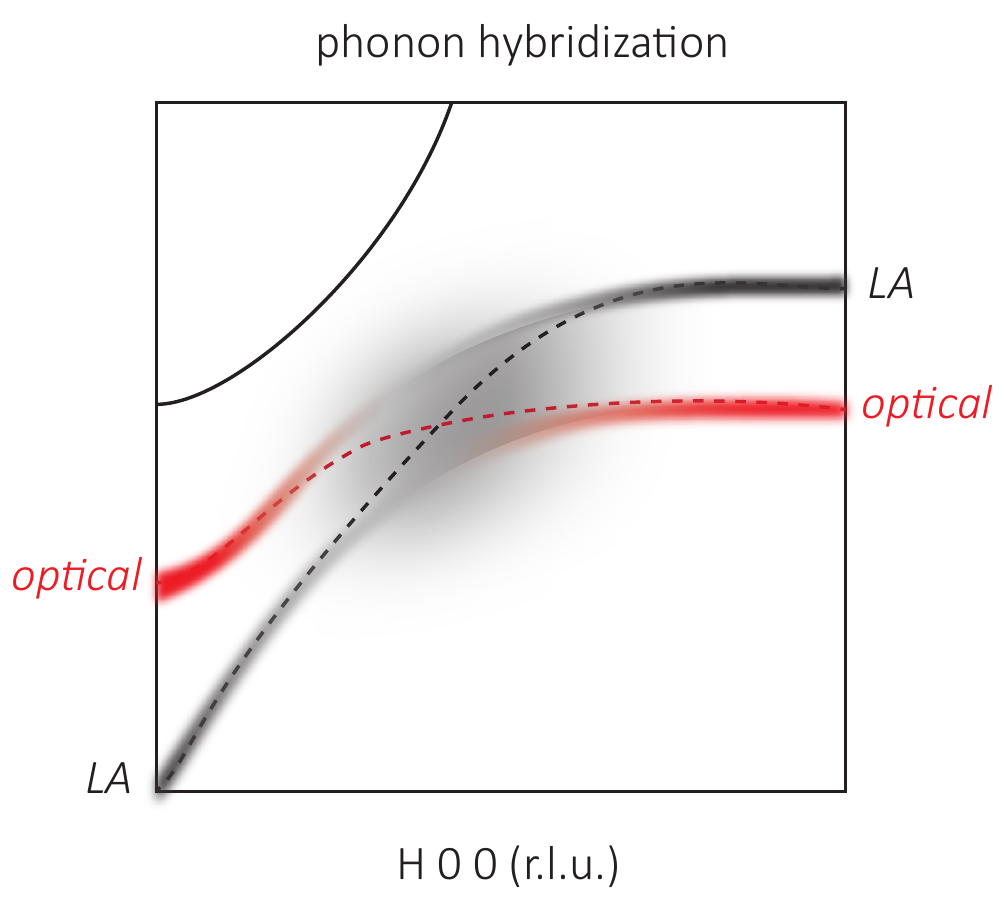}
 	\caption{Intrinsic phonon-phonon hybridization induced broadening. The dashed lines are the non-interacting phonon branches. Red shade indicates the low-energy optical component, and black shade indicates the longitudinal acoustic component. The grey shade indicates the potential hybridization and broadening zone.}
 	\label{fig:figureTHEScenario}
\end{figure}

First, our result reopens the discussion on the scenario where a low-energy forward scattering phonon enhances superconductivity in Bi-2212 system. While previous studies identify the $\sim$ 13-14~meV acoustic phonon to be responsible for the low-energy electron dispersion anomaly in Bi-2212 ~\cite{vishik2010,johnston2012,merritt2017}, our result suggests that the energy position of the electron dispersion anomaly and the small-$q$ requirement of the mode coupling cannot be simultaneously satisfied on the LA/TA branches, mainly due to the energy mismatch at the $q$ of interest. Instead, other higher energy zone center optical phonons are the more likely candidates.

The possible origins of the observed phonon linewidth broadening are discussed schematically in Fig. \ref{fig:figureScenarios} and \ref{fig:figureTHEScenario}. The first explanation requires an existing CDW instability - the low-energy phonon broadens through an interaction with the valence electron CDW and its associated soft mode, as described in panel (a). This is suggested by the evidence in underdoped YBCO and Bi-2212, where the softening and broadening momentum region is extremely narrow, akin to a Kohn anomaly; but the effect maximizes near the superconducting T$_c$ rather than the onset temperature of the CDW ~\cite{blackburn2013,letacon2014,chaix2017,souliou2018}. In this case, the onset of an incommensurate CDW associated with the valence electrons concomitantly triggers broadening and softening in the phonon spectra, displayed as gray regions in Fig.~\ref{fig:figureScenarios}(a). However, as a major caveat to this hypothesis, we find in Bi-2212 that the linewidth broadening does not undergo a change below the onset of the CDW order in UD32. More strongly, the broadening takes place even at the heavily overdoped region $p$ = 0.23, far away from the hitherto known CDW phase region. These combined results rule out this explanation specifically in Bi-2212. We are thus compelled to consider scenarios in which the CDW phenomenon is not the driver of such broadening. 

A second possibility is a non-CDW driven, \textit{q}-dependent electron-phonon coupling for this optical mode, where the mode coupling strength peaks at \textit{q}$_B$ (Fig.~\ref{fig:figureScenarios}(b)). In fact, \textit{q}-dependent electron-phonon coupling is demonstrated to be an important alternative route to realize CDW in correlated systems like NbSe$_2$ ~\cite{weber2011}. Such a property can generally be satisfied for small-\textit{q} scattering phonons, where similar \textit{q}-dependence is expected along both Cu-O and Cu-Cu bond directions ~\cite{johnston2012}. However, as indicated in Fig.~\ref{fig:figure5}(b), such broadening is absent along the (\textit{H} \textit{H}) direction at small \textit{q}, which disfavors the electron-phonon coupling explanation.

 A third, also the most plausible scenario, involves a hybridization of two crossing phonon branches, shown in Fig.~\ref{fig:figureTHEScenario}. As indicated earlier, the LA mode disperses through the low-energy optical mode, intercepting at an intermediate momentum, displayed as a gray area in Fig.~\ref{fig:figureTHEScenario}. The suspected hybridization, moreover, may require the involved optical branch to exhibit anharmonic coupling therefore a quartic term in the local lattice potential. This possibility can be quantitatively evaluated with dynamical lattice calculations.

Therefore, either this phonon broadening is completely independent of the CDW, or the agreement between \textit{q}$_B$ and \textit{q}$_{co}$ is more than mere coincidence. In the latter case, instead of having the CDW as the cause, such low-energy phonon broadening \textit{may} serve as the universal precursory instability to the CDW phase. At this low-energy $E \sim 5$~meV, very little thermal energy is required to significantly populate this phonon at this momentum. When combined with the strong electronic correlation effects (e.g. the pseudogap) in the underdoped region, the electronic instability may be triggered in a concerted way. Indeed, correlation-enhanced electron-lattice interaction has been captured both experimentally and theoretically in underdoped cuprate systems ~\cite{mishchenko2004, YuHe2018_1}. In this coordinated process, the eventual \textit{q}$_{co}$ may slightly shift from the phonon broadening momentum \textit{q}$_B$ to accommodate the electronic correlation effect. This temperature dependent \textit{q}-shift effect has recently been observed in several cuprate systems ~\cite{chaix2017,miao2018}. On the other hand, with the low-energy phonon broadening taking part in the CDW formation, it becomes natural to anticipate a strong pressure effect on the CDW order, as lattice modes are directly perturbed in such a process ~\cite{cyr2015, souliou2018}.

In summary, we have measured high resolution low-energy optical and acoustic phonon dispersions in the high-T$_c$ compound Bi-2212, and achieved quantitative agreement with bulk sound velocity measurements as well as the recent INS measurements. We find the acoustic mode to be unlikely the cause of the 10-15~meV electronic dispersion anomaly in underdoped Bi-2212, while other zone-center optical modes are the more likely candidates.

Importantly, we have observed persistent low-energy longitudinal phonon broadening at \textit{q}$_B \sim 0.28$~r.l.u. along the Cu-O bond direction, similar to the charge order momenta \textit{q}$_{co}$ in many underdoped cuprates. Such broadening has very little temperature dependence up to 240~K. This broadening is likely to happen on a very low-energy optical branch, and may originate from an acoustic-optical phonon crossing. Combining non-resonant and resonant x-ray scattering techniques, we find no evidence for CDW in the heavily overdoped Bi-2212. These suggest that if the low-energy phonon broadening were to be related to the CDW in the underdoped cuprates at all, in lieu of being induced by the CDW, it actually may provide the precursory instability to the CDW formation, which is eventually triggered when combined with the much stronger electronic correlations in the underdoped phase region.

\section{Acknowledgement}
The authors wish to thank S. Johnston, T. P. Devereaux, S. C. D. Pemmaraju and S. Sinha for helpful discussions. The work at the University of California, Berkeley and Lawrence Berkeley National Laboratory is supported by the Office of Science, Office of Basic Energy Sciences, Materials Sciences and Engineering Division, of the U.S. Department of Energy (DOE) under Contract No. DE-AC02-05CH11231 within the Quantum Materials Program (KC2202). The works at Stanford University and Stanford Synchrotron Radiation Lightsource, SLAC National Accelerator Laboratory, are supported by the U.S. Department of Energy, Office of Science, Office of Basic Energy Sciences under Contract No. DE-AC02-76SF00515.  This research used resources of the Advanced Photon Source, a U.S. Department of Energy (DOE) Office of Science User Facility operated for the DOE Office of Science by Argonne National Laboratory under Contract No. DE-AC02-06CH11357.Part of this work is performed at the Stanford Nano Shared Facilities (SNSF), supported by the National Science Foundation under award ECCS-1542152. M.Y. acknowledges the L'Or\'eal For Women in Science Fellowship for support.

\bibliographystyle{apsrev4-1}

\end{document}